\documentclass[11pt,twoside,dvips]{article}
\usepackage{asp2010}

\resetcounters

\newcommand{\Kelvin}{\mathrm{K}}

\begin{document}

\title{Simulations of Accretion onto Magnetized Stars: Results of 3D MHD
Simulations and 3D Radiative Transfer}
\author{Marina~Romanova,$^1$ Ryuichi Kurosawa,$^2$}

\begin{abstract}

We discuss the results of modelling of young magnetized stars,
where matter flow is calculated using the three-dimensional (3D)
magneto-hydrodynamic (MHD) \textit{Cubed Sphere} code, and the
spectra are calculated using the 3D Monte Carlo radiative transfer
code \textit{TORUS}. Two examples of modelling are shown: (1)
accretion onto stars in stable and unstable regimes, and (2)
accretion to a young star V2129 Oph, modelled with realistic
parameters.

\end{abstract}

\section{Introduction}
\label{sec:introduction}

The low-mass pre-main sequence solar-type stars evolve through
different stages. Many of them are at the stage of a classical T
Tauri star (CTTS),  where the star becomes visible, but is still
surrounded by a protoplanetary disk (e.g., \citealt{bouv07}).
Observations show that CTTSs usually have a strong, dynamically
important magnetic field.  In these stars, the accretion disc is
truncated by the magnetosphere, and the magnetic field governs the
matter flow (e.g., \citealt{prin72,ghos79}). The photometric and
spectral variabilities of these stars are determined by the
patterns of matter flow through the magnetosphere and by the
shapes and positions of the hot spots.  This problem is
three-dimensional, so that the matter flow should be studied in
global 3D MHD simulations, while photometry and spectra should be
calculated using 3D radiative transfer approaches.

\section{Numerical approaches}
\label{sec: numerical approaches}

 We perform global simulations of
matter flow around magnetized young stars using the 3D MHD
\textit{Cubed Sphere} code, and afterwards use the obtained
results for the calculation of spectra using the 3D radiative
transfer code \textit{TORUS}.

\medskip

\noindent\textbf{2.1. 3D MHD Simulations with \textit{Cubed
Sphere} code} \label{sec:cubed sphere}

\smallskip

We use the second-order Godunov-type three-dimensional (3D) MHD
code developed by our group \citep{kold02}. It has many specific
features which are oriented towards the efficient calculation of
accretion onto a star with a tilted dipole or a more complex
magnetic field: (1) the magnetic field $\bf B$ is decomposed into
the ``main'' dipole component of the star, ${\bf B}_0$, and the
component ${\bf B}_1$ induced by currents in the disc and the
corona \citep{tana94}; (2) the MHD equations are written in a
reference frame rotating with the star; and (3) the numerical
method uses the ``cubed sphere" grid.  The grid on the surface of
the sphere consists of six sectors, with the grid in each sector
being topologically equivalent to the grid on a face of a cube
(e.g., \citealt{ronc96}). We use a Godunov-type numerical scheme
similar to the one described by \citet{pow99} and perform
simulations in three dimensions. The full set of equations is the
following:

\begin{figure*}
\centering
\includegraphics[width=4.1cm]{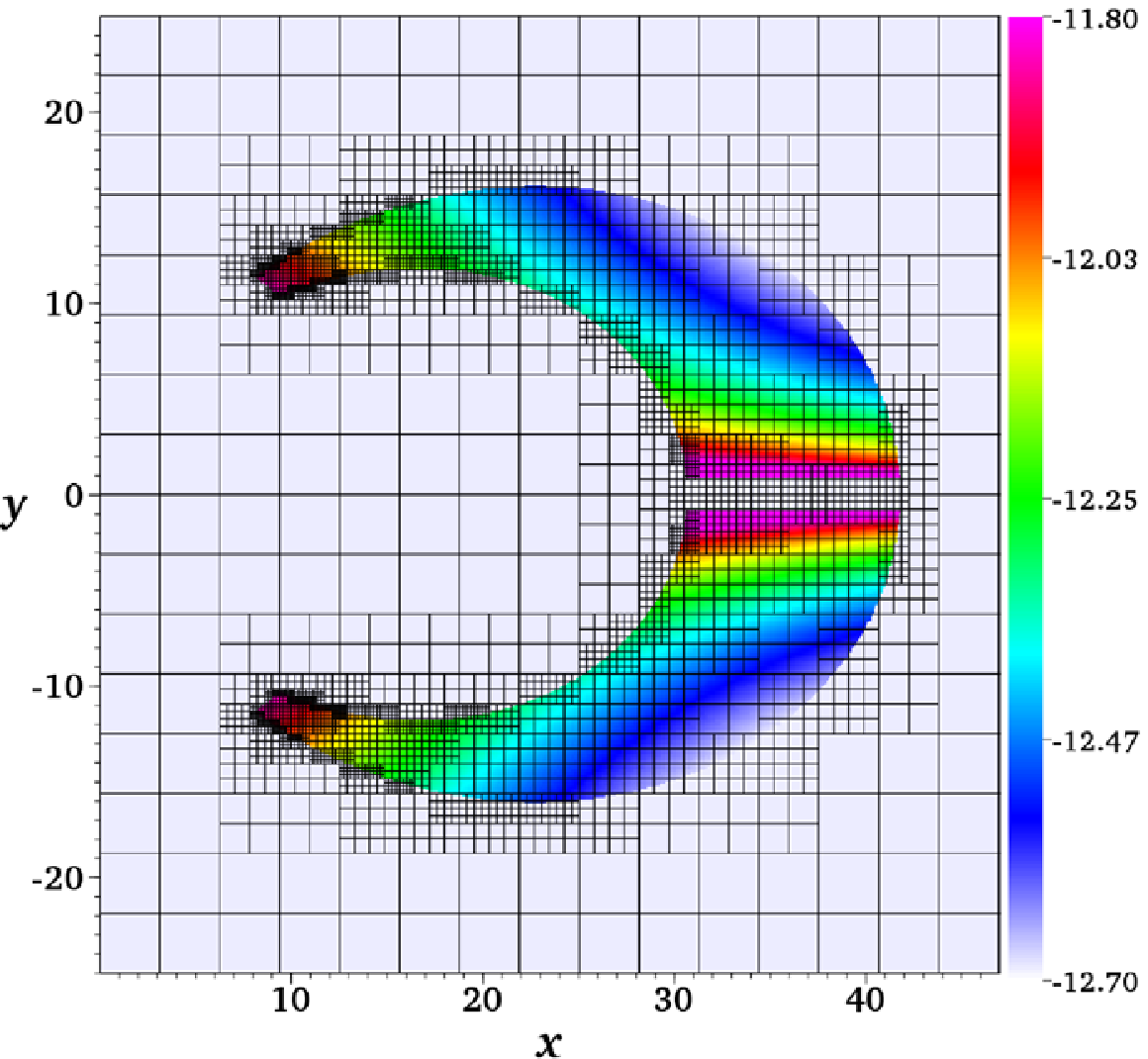}
\includegraphics[width=4.3cm]{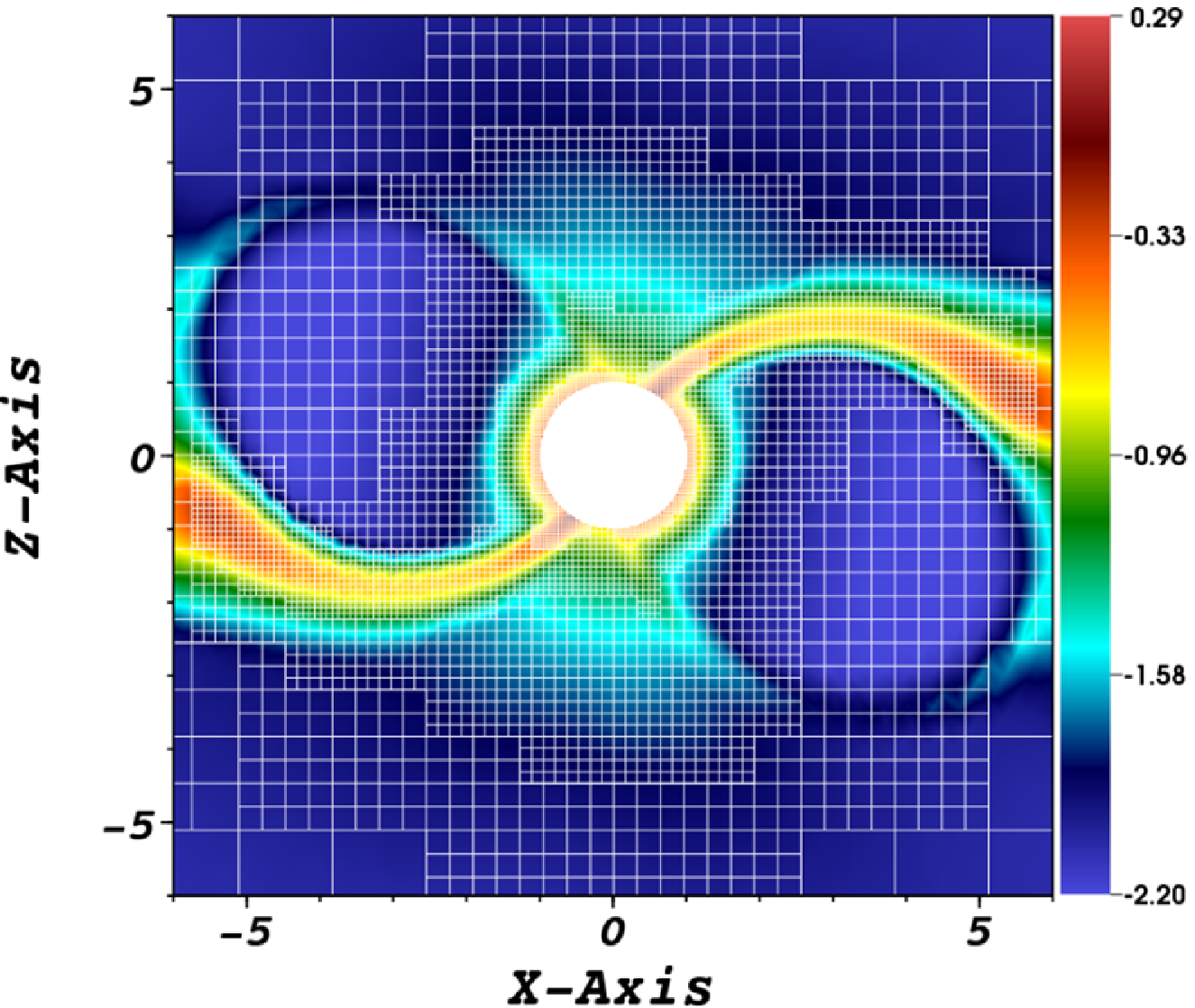}
\vspace{-0.5cm}\caption{\textit{Left Panel:} An example of the
restructuring grid of radiative transfer code \textit{TORUS} in
the case where the funnel stream density and other parameters are
determined by the analytical formula of Hartman et al. (1994)
(from Kurosawa et al. 2004).  \textit{Right panel:} same as left
panel, but for the case where the funnel flow was obtained in 3D
MHD simulations (from \citealt{kurosawa08}).} \label{funnel-grid}
\end{figure*}

$$ {{\partial \rho}/{\partial t}} + {\bf {\nabla}}\cdot (\rho{\bf v}) = 0, $$
 $$ {\partial (\rho{\bf  v})}/{\partial t} + {\bf
{\nabla}}\cdot {T} = \rho {\bf g} + 2\rho ~{\bf v}\times{\bf \Omega} - \rho~ {\bf
\Omega}\times ({\bf\Omega}\times{\bf R}), $$ $$ {\partial (\rho S)}/{\partial t} +
{\bf {\nabla}}\cdot (\rho S {\bf v}) = 0~, $$
$$ {\partial {\bf B}}/{\partial t} = {\bf \nabla
\times} ({\bf v}\times{\bf B}), $$ where  ${\bf v}$ is the
velocity of plasma in the rotating frame,  ${\bf B}$ is the
magnetic field, and $S$ is the specific entropy.
 $T$ is the stress tensor with components $T_{ik} \equiv
p\delta_{ik} +\rho v_i v_k+ ({B^2}\delta_{ik}/2-B_iB_k)/4\pi$, and $p$ is the gas
pressure.

    All variables are evaluated at the
centers of the cells, and all vector variables are expressed in
terms of their Cartesian components. Finite difference equations
are then written for the variables. The finite difference scheme
of Godunov's type has the form: $ {{{\cal U}^{p+1}-{\cal
U}^p}/{\Delta t}} V + \sum_{m=1..6} s_m {\cal F}_m = {\cal Q}~. $
  Here,
${\cal U}=\left \{\rho,~ \rho {\bf v},
 {\bf B},
~\rho S\right \} $  is the ``vector" of the densities of conserved
variables; ${\cal F}_m$ is the ``vector" of flux densities normal
to the face ``$m$" of the grid cell, $s_m$ is the area of the face
``$m$", $V$ is the volume of the cell,  $\cal Q$ is the intensity
of sources in the cell, and $\Delta t$ is the time step. To
calculate the flux densities ${\cal F}_m$, an approximate Riemann
solver is used, analogous to the one described by \citet{pow99}
(see also \citealt{kuli01}). The grid resolution is either
$N_x=N_y=51$ or $61$ in each of the 6 blocks of the cube.
 The number of grid cells in the radial direction is $N_r=150-180$.

\begin{figure*}
  \begin{center}
    \begin{tabular}{cc}
             \includegraphics[clip,height=0.35\textwidth]{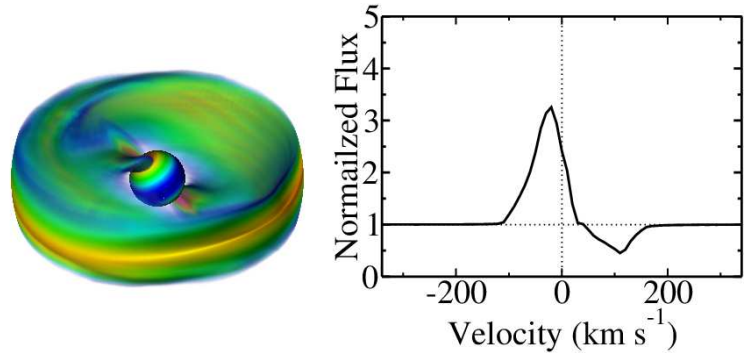}
    \end{tabular}
  \end{center}
\vspace{-0.8cm} \caption{An example of accretion in the stable
regime, where matter accretes in two ordered funnel streams (see
density distribution in the left panel). The right panel shows the
spectrum in H$\delta$ hydrogen spectral line. The red-shifted
absorption appears two times per period when the funnel stream
crosses the line-of-sight. From \citet{KurosawaRomanova2013}.}
\label{stable-rho-hdelta}
\end{figure*}

\medskip

\noindent\textbf{2.2. 3D Radiative Transfer Code \textit{TORUS}}
\label{sub:mhd-rad:rad}

\smallskip
 For the calculations of hydrogen emission line profiles from the
 matter flow in the MHD simulations, we use the radiative transfer
 code \textit{TORUS} (e.g.~\citealt{harries2000, harries2011,
   kurosawa06, kurosawa11, kurosawa12}). In particular, the numerical
 method used in the current work is essentially identical to that in
 \citet{kurosawa11}.

The basic steps for computing the line variability are as follows:
(1)~mapping the MHD simulation data onto the radiative transfer
grid, (2)~source function calculations, and (3)~observed line
profile calculations as a function of rotational phase. In
step~(1), we use an adaptive mesh refinement (AMR) which allows
for the accurate mapping of the original MHD simulation data onto
the radiative transfer grid (Fig.~\ref{funnel-grid}). In step~(2),
we use a method similar to that of \citet{klein78} (see also
\citealt{rybicki78}; \citealt*{hartmann94}) in which the Sobolev
approximation (e.g.~\citealt{sobolev1957,castor1970}) is applied.

The Sobolev approximation works when the velocity gradient in the
medium is large, such that a line center photon does not interact
with the surrounding medium due to the Doppler effect, except for
sharp resonance zones along a given direction. This essentially
reduces the computation of the radiation field to a local problem,
as opposed to a global problem. Normally, the radiation field at a
given point in the medium depends on the radiation fields at all
the points in the medium; hence, evaluating radiation fields in
two or three dimensions is a computationally challenging problem.
The use of the approximation significantly reduces the
computational time, thus allowing us to make a multi-dimensional
problem feasible. For example, the line profile averaged mean
intensity at a given point can be expressed as (e.g.,
\citealt{rybicki78})
\begin{equation}
  \bar{J}=\left(1-\beta\right)S_{l} + I_{\mathrm{c}}\beta_{\mathrm{c}}
  \label{eq:mean-intensity}
\end{equation}
where $S_{l}$ is the line source function and $I_{\mathrm{c}}$ is
the intensity from the continuum radiation source (assuming no
limb-darkening). The term $S_{l}$ is local, i.e., it depends only
on the conditions of local gas (e.g.~level populations and so on),
while $I_{\mathrm{c}}$ comes from a boundary condition. Further,
the quantities $\beta$ and $\beta_{\mathrm{c}}$ are the angle
averaged escape probabilities of a photon from the point where
$\bar{J}$ is evaluated, and they depend on the local quantities:
the Sobolev optical depth $\tau_{\mathrm{s}}$ and the direction of
photon propagation $\mathbf{n}$. The former can be expressed as
\begin{equation}
  \tau_{\mathrm{s}}=\frac{c\,\chi_{l}}{\nu_{l}}\left|\frac{dv_{\mathbf{n}}}{dL}\right|^{-1},
  \label{eq:tau-sobolev}
\end{equation}
where $c$, $\chi_{l}$ and $\nu_{l}$ are the speed of light, line
opacity, and line frequency, respectively. The last term
$dv_{\mathbf{n}}/dL$ is the velocity gradient along the line
element $dL$ in the direction of the photon propagation
$\mathbf{n}$.

The populations of the bound states of hydrogen are assumed to be
in statistical equilibrium, and the continuum sources are the sum
of radiations from the stellar photosphere and the hot spots
formed by the funnel accretion streams falling onto the stellar
surface. Our model hydrogen atom consists of 20 bound and
continuum states. For the photospheric contribution to the
continuum flux, we adopt the effective temperature of the
photosphere $T_{\mathrm{ph}}=4000\,\Kelvin$ and the surface
gravity $\log g_{*}=3.5$ (cgs), and use the model atmosphere of
\citet{kurucz1979}.  The sizes and shapes of the hot spots are
determined by the local energy flux on the stellar surface
(\citealt{RomanovaEtAl2004}).

In step~(3), the line profiles are computed using the source
function computed in step~(2). The observed flux at each frequency
point in the line profiles is computed using the cylindrical
coordinate system, with its symmetry axis pointing towards the
observer. The viewing angles of the system (the central star and
the surrounding gas) are adjusted according to the rotational
phase of the star and the inclination angle of the system for each
time-slice of the MHD simulations.

\begin{figure*}
  \begin{center}
    \begin{tabular}{cc}
             \includegraphics[clip,height=0.33\textwidth]{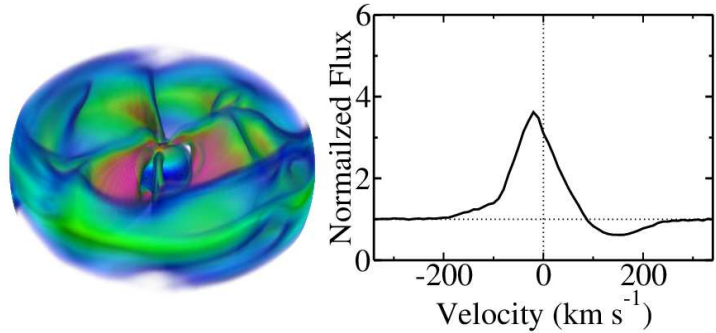}
    \end{tabular}
  \end{center}
\vspace{-0.8cm} \caption{Same as in Fig. \ref{stable-rho-hdelta},
but for the case of unstable accretion, where matter accretes in
several unstable tongues and the red-shifted absorption varies
irregularly,  but is always present in the spectrum, because there
are always tongues on the line-of-sight. From
\citet{KurosawaRomanova2013}.} \label{unstable-rho-hdelta}
\end{figure*}

\section{Examples of modelling}

Here, we give two examples of modelling of young stars using 3D
MHD + 3D radiative transfer approach.

\subsection{Spectral diagnostics of stable and unstable regimes of accretion}

A magnetized star with a dipole magnetic field may accrete in
either stable or unstable regime
\citep{RomanovaEtAl2008,KulkarniRomanova2008}. In the stable
regime, matter flows above the magnetosphere in two ordered funnel
streams, and the two hot spots on the surface of the star provide
a nearly sinusoidal pattern of variability. In the unstable
regime, matter penetrates between the magnetic field lines due to
the magnetic Rayleigh-Taylor instability (e.g.,
\citealt{AronsLea1976}). Simulations show that matter may accrete
to the star in several unstable ``tongues", which form irregular
hot spots on the surface of the star, and the light-curve may be
irregular. However, there are other possible causes for the
irregular light-curves, such as frequent stellar magnetic flares,
analogous to the solar flares, or accretion from a turbulent disk.
Therefore, the irregular photometric light-curve alone is not a
proof of unstable accretion. That is why we performed global 3D
simulations of stable and unstable regimes of accretion, and used
the results of simulations for the calculation of spectra in the
Hydrogen spectral lines \citep{KurosawaRomanova2013}. We chose 25
moments in time per rotational phase of the star, mapped the
calculated values of matter flow (density, velocity, scaled
temperature) to the AMR grid of the \textit{TORUS} code and
calculated the spectra for three rotations of the star (75 total).
In the case of stable accretion, we observed that the spectrum has
a typical red-shifted absorption when the funnel stream is on the
line-of-sight between the star and the observer (see Fig.
\ref{stable-rho-hdelta}). It is absent when the funnel streams are
away from the observer. In the unstable regime, matter flows to
the star in several unstable tongues, and we observed red
absorption all the time (see Fig. \ref{unstable-rho-hdelta}). In
the unstable regime, the red-shifted absorption varies
irregularly. This analysis may help to distinguish young stars
accreting in the unstable regime. Many young, classical T Tauri
stars (CTTSs) show irregular variability \citep{AlencarEtAl2010}
on the time-scales corresponding to unstable accretion. Our
analysis may help understand whether unstable accretion is
responsible for this variability.

\begin{figure*}
  \begin{center}
    \begin{tabular}{cc}
\includegraphics[clip,height=0.25\textwidth]{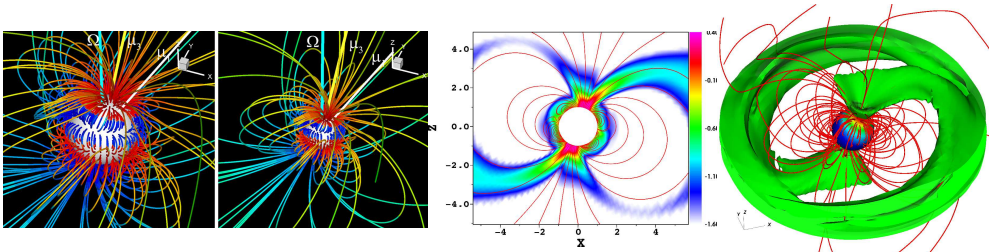}
    \end{tabular}
  \end{center}
\vspace{-0.9cm} \caption{\textit{Left 2 panels:} The magnetic
field lines in the model of V2129 Oph, where the dipole and
octupole components dominate. \textit{Right two panels:} 2D and 3D
views of matter flow around the modeled star V2129 Oph. From
Romanova et al. (2011).} \label{v2129-2d-3d}
\end{figure*}

\subsection{Modelling accretion onto V2129 Oph}

\begin{figure*}
  \begin{center}
    \begin{tabular}{cc}
             \includegraphics[clip,height=0.35\textwidth]{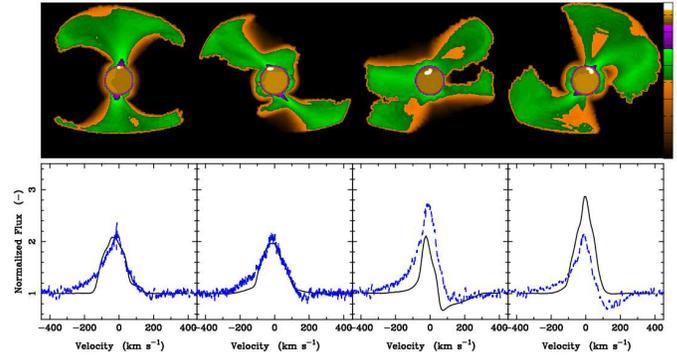}
    \end{tabular}
  \end{center}
\vspace{-0.8cm} \caption{\textit{Top panels:} Emissivity of the
funnel flow calculated in the H$\beta$ spectral line.
\textit{Bottom panels:} Comparison of the observed spectrum in
H$\beta$ line (blue line) with the modelled spectrum (black line).
From Alencar et al. (2012).} \label{v2129oph-hbeta}
\end{figure*}

Recently, the surface distribution of the magnetic field has been
measured for several CTTSs. We chose one of these stars, V2129
Oph, and took the parameters of this star derived from
observations: $M_*=1.35 M_\odot$, $R_*=2.1 R_\odot$, and period
$P_*\approx 6.5$ days. The magnetic field of this star is
dominated by the dipole component of
 $B_{dip}\approx 0.9$kG and octupole component of
$B_{oct}\approx 2.1$kG, tilted at
small angles about the rotational axis \citep{DonatiEtAl2011}. We
took the magnetic field configuration consisting of the dipole and
octupole, and developed the MHD model based on these observed
parameters of the star (see also \citealt{RomanovaEtAl2011}). Fig.
\ref{v2129-2d-3d} shows the results of simulations. We mapped the
results of simulations to the AMR grid of the \textit{TORUS} code,
and calculated the spectrum in different hydrogen lines. Fig.
\ref{v2129oph-hbeta} shows that the observed spectrum is in good
agreement with the modelled spectrum \citep{AlencarEtAl2012}. This
is an exciting example where a star with realistic parameters has
been modelled in 3D MHD + 3D radiative transfer simulations, and
the result has been compared with observations.

\acknowledgements  We thank the organizers for a very interesting
meeting. Resources supporting this work were provided by the NASA
High-End Computing (HEC) Program through the NASA Advanced
Supercomputing (NAS) Division and the NASA Center for
Computational Sciences (NCCS). The research was supported by NASA
grant NNX11AF33G and NSF grant AST-1211318.

\end{document}